# Comment: Bibliometrics in the Context of the UK Research Assessment Exercise

**Bernard W. Silverman**

*Abstract.* Research funding and reputation in the UK have, for over two decades, been increasingly dependent on a regular peer-review of all UK departments. This is to move to a system more based on bibliometrics. Assessment exercises of this kind influence the behavior of institutions, departments and individuals, and therefore bibliometrics will have effects beyond simple measurement.

In the United Kingdom's Research Assessment Exercise (RAE), every university may submit its research in every discipline for assessment. On this assessment rests a considerable amount of funding; indeed a number of universities, leading "research universities" in American nomenclature, gain more from this source of research funding than from government funding for teaching. Within broad subject bands, the Higher Education Funding Council for England funds teaching on a flat rate per student. So the amount of funding a student of Mathematics attracts is the same whichever university they attend. On the other hand, funding for research is selective: those departments which fare well on the Research Assessment Exercise receive more funding as a result. This is in addition to any income from grants and grant overheads.

The RAE and its predecessors have been running for over two decades, and have always been based on peer review, though numerical data on student numbers and grant income also have some input into the assessment. However, it is proposed that "metrics," which include so-called bibliometric data, will be the main part of the system which will soon succeed the RAE, though it is probable that in mathematical subjects, peer review will continue to play a considerable part. The details have yet to be worked out.

*B. W. Silverman is Master, St Peter's College, Oxford OX1 2DL, United Kingdom e-mail: bernard.silverman@spc.ox.ac.uk.*

In the 2008 RAE, I was chair of the committee which reviewed Probability, Statistics and the more mathematical aspects of Operational Research. The committee's experience of conducting the assessment as a whole strengthened our view that peer review must be at the core of any future assessment of research in our area. Reliance on bibliometric and purely quantitative methods of assessment would, in our unanimous view, introduce serious biases, both into the assessment process and, perhaps more seriously, into the behavior of institutions and of individual researchers, to the detriment of the very research which the exercise is intended to support.

It is important to stress the effect of any system on the behavior of institutions. The current peer-review RAE has had clear effects on institutional behavior, some of them certainly positive, some of them perhaps less so. For example, the RAE gives explicit advantages to new entrants to the profession; those entering in the last few years are allowed to submit a smaller corpus of work for assessment, and there is also credit given within the peer review system for a subjective assessment of the general vitality of the department. Of the approximately 400 research-active faculty declared to the statistics panel in the 2008 RAE, about a quarter were new entrants since 2001, and the RAE has certainly given an impetus to this new recruitment, as it also does to the mobility of leading researchers between institutions. On a more negative note, the fixed date of the assessment encourages a "boom-bust" mentality, where some institutions hire in considerable numbers of new faculty in the period leading up to the census date; to make up for this extra expenditure, during the period after the census date there is something of a moratorium on appointments. The consideration of grant income in the RAE gives extra gearing to the pressure on faculty to pursue grant-supported research rather than to work in a more individual fashion.

There can be little doubt that a stronger emphasis on bibliometrics (and other "metrics") in assessment exercises will affect institutional behavior, especially in systems where assessment results have both reputational and fiscal impact. Because individuals are sensitive to institutional pressures, they too will modify their behavior in response. For example, it is probably the case that there is a high correlation between $h$-index (say) and perceived quality and reputation of researchers. Similarly, highly-cited papers are almost always influential and important (though the converse is not necessarily true). However, basing judgment of individuals or departments on citation count rather than some assessment of underlying quality would have the obvious perverse consequences. Perhaps the obvious analogy would be with a system that counts publications: of course there is some correlation between the overall quality of a researcher's work and the number of papers she or he publishes, but the "publish or perish" mentality engendered by simple paper-counting militates against the careful and thoughtful researcher who only writes papers when they feel they have something very serious to say, or—worse still—writes books rather than papers. Perhaps the bibliometric version is "be cited or benighted"?

One of the arguments the UK university funding agencies used initially in favor of bibliometrics was that, when aggregated over whole universities, the results of "metrics-based" assessments were very highly correlated with peer-review assessments. As statisticians, we should be well placed to refute the fallacy of this argument. It makes complete sense that a strong university will have more than its fair share of highly-cited researchers right across the range of disciplines. Any errors and biases will to some extent average out. But disaggregation down to departments, and even individuals, encourages the elimination, or downgrading, of disciplines and sub-disciplines which do not generate large amounts of citations. Within disciplines, there is a risk of undervaluing individuals whose work is deeply influential in ways that do not show up in short-term citation counts. And many individual researchers would no doubt bow to perceived pressure to be "cited or benighted."

If citation counts are unreasonable, the use of impact factors seems almost indefensible. Assigning a notion of quality to a paper on the grounds of the impact factor of a journal is like assigning a notion of wealth to an individual on the basis of the average GDP of their home country. Many children growing up in England in the 1950s were under the impression that all Americans were wealthy! Of course, if one knows about the refereeing standards of a leading journal, it may, or may not, be reasonable to suppose that if a paper has passed these standards it has a good chance of being of high quality, but that is a very different thing from assessing the journal by an impact factor.

In conclusion, I would very strongly support the underlying thesis of the paper: citation statistics, impact factors, the whole paraphernalia of bibliometrics may in some circumstances be a useful servant to us in our research. But they are a very poor master indeed.